\RequirePackage{fix-cm}
\documentclass[twocolumn]{svjour3}          
\smartqed  
\usepackage{amsmath,amssymb,amsfonts}
\usepackage{graphicx}
\usepackage{xcolor}
\usepackage{url}
\usepackage{subfigure}
\usepackage{tcolorbox}
\usepackage{multirow}
\usepackage{listings}
\usepackage{amssymb,amsfonts}
\usepackage{pifont}
\usepackage{xcolor,colortbl} 
\usepackage{listings}
\usepackage{color,hyperref}
\usepackage{marvosym}

\lstdefinestyle{JavaStyle} {
  backgroundcolor=\color{white},   
  commentstyle=\color{mygreen}, 
  breakatwhitespace=false,
  keywordstyle=\color{violet},
  language=Java,
  stringstyle=\color{blue},
  basicstyle=\scriptsize\ttfamily,
  showstringspaces=false }

\definecolor{mygreen}{rgb}{0,0.6,0}
\definecolor{mygray}{rgb}{0.95,0.95,0.95}
\definecolor{myred}{rgb}{0.5,0,0}

\newcommand{\cmark}{\ding{51}}
\newcommand{\xmark}{\ding{55}}

\usepackage{xspace}
\newcommand*{\ie}{i.e.,\@\xspace}
\newcommand*{\eg}{e.g.,\@\xspace}

\newcommand*{\GH}{GitHub\@\xspace}
\newcommand*{\EP}{EvoPlan\@\xspace}
\makeatletter
\newcommand*{\etc}{%
	\@ifnextchar{.}%
	{etc}%
	{etc.\@\xspace}%
}
\makeatother
\newcommand*{\etal}{\emph{et~al.}\@\xspace}
\newcommand\revised[1]{\textcolor{black}{#1}}

\newcommand{\rqfirst}{\textbf{RQ$_1$}: \emph{How effective is \EP in terms of prediction accuracy?}} 
\newcommand{\rqsecond}{\textbf{RQ$_2$}: \emph{Is there any correlation between the \GH issues and the popularity of a certain migration path?}}
\newcommand{\rqthird}{\textbf{RQ$_3$}: \emph{Is \EP able to provide consistent recommendations in reasonable time?}}

\definecolor{Crimson}{rgb}{0.86,0.07,0.23}
\definecolor{Gold}{rgb}{0.98,0.65,0.0}

\definecolor{darkgray}{gray}{0.78}
\definecolor{lightgray}{gray}{0.85}
\definecolor{verylightgray}{gray}{0.95}

\begin{document}

\title{Recommending Upgrade Plans for Third-party Libraries with Migration Graphs}

\title{\EP: A Recommender System using Migration Graphs to Provide Upgrade Plans for Third-party Libraries}

\title{Providing Upgrade Plans for Third-party Libraries: A Recommender System using Migration Graphs}


\author{Riccardo Rubei \and Davide Di Ruscio \and Claudio Di Sipio \and Juri Di Rocco 
\and Phuong T. Nguyen}

\institute{
	Riccardo Rubei \at
	Universit\`a degli studi dell'Aquila, Italy \\ 
	\email{\href{mailto:riccardo.rubei@graduate.univaq.it}{riccardo.rubei@graduate.univaq.it}}
	\and
	\Letter~Davide Di Ruscio \at
	Universit\`a degli studi dell'Aquila, Italy \\ 
	\email{\href{mailto:davide.diruscio@univaq.it}{davide.diruscio@univaq.it}}
	\and
	Claudio Di Sipio \at
	Universit\`a degli studi dell'Aquila, Italy \\ 
	\email{\href{mailto:claudio.disipio@graduate.univaq.it}{claudio.disipio@graduate.univaq.it}}
	\and
	Juri Di Rocco \at
	Universit\`a degli studi dell'Aquila, Italy \\ 
	\email{\href{mailto:juri.dirocco@univaq.it}{juri.dirocco@univaq.it}}	
	\and
	Phuong T. Nguyen \at
	Universit\`a degli studi dell'Aquila, Italy \\ 
	\email{\href{mailto:phuong.nguyen@univaq.it}{phuong.nguyen@univaq.it}}           
}
\maketitle

\begin{abstract}
		
		During the development of a software project, \revised{developers often need to upgrade third-party libraries (TPLs), aiming to keep their code up-to-date with the newest 
		functionalities offered by the used libraries.} In most cases,  upgrading used TPLs is a complex and error-prone activity that must be carefully carried out 
		to limit the ripple effects on the software project that depends on the 
		libraries being upgraded. 		
		In this paper, we propose \EP 
		as a novel approach to the recommendation of different upgrade plans given a pair of library-version as input. In particular, among the different paths that can be possibly followed to upgrade the current library 
		version to the desired updated one, \revised{\EP is able to suggest the plan that can potentially minimize the efforts being needed to migrate the code of the clients from the library's current release to the target one.} The	approach has been evaluated on a curated dataset using conventional metrics 
		used in Information Retrieval, \ie precision, recall, and F-measure. The 
		experimental results show that EvoPlan obtains an encouraging prediction 
		performance considering two different criteria in the plan specification, \ie 
		the popularity of migration paths and the number of open and closed issues in 
		\GH for those projects that have already followed the recommended 
		migration paths.
		
		
%
\end{abstract}

		\section{Introduction} 
		\label{sec:Introduction}


When dealing with certain coding tasks, developers usually 
make use of third-party libraries (TPLs) that provide the 
desired functionalities. Third-party libraries offer a wide range of 
operations, \eg database management, file utilities, Website 
connection, to name a few. Their reuse
allows developers to exploit a well-founded 
infrastructure, without reinventing the wheel, which 
eventually helps save time as well as increase productivity. 
\revised{However, TPLs evolve over the course of time, and API functions can either be added or removed, aiming to make the library become more efficient/effective, as well as to fix security issues.} 
\revised{Upgrading clients' code from a library release to a newer one can be a daunting and time-consuming task, especially when the APIs being upgraded introduce breaking changes that make the client fail to compile or introduce  behavioral changes into it \cite{7884616}. Thus, managing TPLs and keeping them up-to-date becomes a critical practice to minimize the technical debt~\cite{avgeriou_et_al:DR:2016:6693}.}

%

\revised{In order to upgrade a client $C$ from a starting library version $l_{v_i}$ to a target one $l_{v_t}$, the developer needs to 
understand both versions' documentation deeply, as well as to choose the right matching between corresponding methods.}
Things become even more complicated when several subsequent 
versions of the library of interest $l$ have been released 
from $v_i$  to  $v_t$. 
In such cases, developers who want to reduce the technical 
debts, which have been accumulated due to libraries 
that have not been upgraded yet, have first to decide the 
\textit{upgrade plan} that has to be applied, i.e., how to go 
from $l_{v_i}$  to  $l_{v_t}$ since many possible paths might 
be followed. In such cases, it is essential to have proper 
machinery to assist developers in choosing suitable upgrade 
plans to potentially reduce the efforts that are needed to migrate the 
client project  $C$  under development. It is possible to minimize migration efforts by identifying upgrade plans, which similar projects have already performed, and thus, by relying on the experiences of already upgraded clients. In this way, developers have the availability of supporting material, \eg documentation, and snippets of code examples that can be exploited during the migration phases.


In the context of open-source software, developing new systems by reusing existing components raises relevant challenges in: \textit{(i)} searching for relevant modules;  and \textit{(ii)} adapting the selected components to meet some pre-defined requirements. To this end, recommender systems in software engineering have been developed to support developers in their daily tasks~\cite{robillard_recommendation_2014,di_rocco_development_2021}. Such systems have gained traction in recent years as they are able to provide developers with a wide range of useful items, including code snippets~\cite{Nguyen:2019:FRS:3339505.3339636}, tags/topics \cite{10.1145/3382494.3410690,10.1145/3383219.3383227}, third-party libraries \cite{Nguyen:2019:JSS:CrossRec}, documentation~\cite{ponzanelli_prompter:_2016,RUBEI2020106367}, to mention but a few. \revised{In the CROSSMINER project~\cite{di_rocco_development_2021}, we conceptualized various techniques and tools for extracting knowledge from open source components to provide tailored recommendations to developers, helping them complete their current development tasks.}

\revised{In this work, we propose \EP, a recommender system to provide 
upgrade plans for TPLs. 
By exploiting the experience of other projects that have already performed similar 
upgrades and migrations, \EP recommends the plan that should be considered to upgrade from the current library 
version to the desired one. A graph-based representation is inferred by analyzing \GH repositories and their \emph{pom.xml} files.} During this phase, 
\EP assigns weights representing the number of client projects that have already performed a specific upgrade. 
Afterwards, the system employs a shortest-path algorithm to minimize the 
number of upgrade steps considering such weights. It eventually retrieves
multiple upgrade plans to the user with the target version as well as all the intermediate passages.

%
To the best of our knowledge, there exist no tools that provide this type of recommendations. Thus, we cannot compare \EP with any baselines but evaluate it by using metrics commonly used in information retrieval 
applications, \ie precision, recall, and F-measure. 
Furthermore, we also evaluate the correlation between \GH \footnote{\url{https://github.com/}} issues data
and the suggested upgrade plans.
In this sense, our work has the following contributions: 


\begin{itemize}
	\item \emph{Gathering and storing of migration data}: Using \textsc{Neo4j} Java Driver,\footnote{\url{https://github.com/neo4j/neo4j-java-driver}} \EP stores the extracted data in a persistent and flexible data structure;
	\item \emph{Recommendation of an upgrade plan list}: Considering the number of clients, \EP suggests the most common upgrade plans that are compliant with those that have been accepted by the developers community at large;
	\item \emph{Modularity and flexible architecture}: The proposed system can be seen as both an external module integrable into other approaches and a completely stand-alone tool that can be customized by end users; 
	\item \emph{Automated evaluation and replication package availability}: \revised{The performance of \EP has been evaluated by employing the widely used ten-fold cross-validation technique. Last but not least, we make the \EP replication package available online to facilitate future research.}\footnote{\url{https://github.com/MDEGroup/EvoPlan}} 
\end{itemize}

%
%
%
%

The paper is structured as follows. Section 
\ref{sec:Background} presents a motivating example and existing migration tools 
in the literature. Furthermore, in this section we also highlight the open 
challenges in the domain. Section \ref{sec:ProposedApproach} introduces \EP, 
\revised{the proposed approach to the recommendation of third-party library upgrades.} In Section 
\ref{sec:Study}, we present the performed evaluation process. The results obtained from the empirical 
evaluation are presented in Section \ref{sec:Results} together with possible threats to validity. The related work is reviewed in Section \ref{sec:RelatedWorks}. Finally, we conclude the paper and envisage future work 
in Section~\ref{sec:Conclusion}.

	\section{Motivations and Background}
	\label{sec:Background}

TPLs offer several tailored functionalities, 
and invoking them allows developers to make use of a well-founded infrastructure, without needing to re-implementing from scratch~\cite{Nguyen:2019:JSS:CrossRec}. Eventually, this helps save time as well as increase productivity. However, as libraries evolve over the course of time, it is necessary to have a proper plan to migrate them once they have been updated. 
So far, various attempts have been made to tackle this issue. 
In this section, we introduce two motivating examples, and recall some notable relevant work as a base for further presentation.

\subsection{Explanatory examples} \label{sec:exp}
This section discusses two real-world situations that developers must cope with during the TLPs migration task, \ie code refactoring and vulnerable dependencies handling. In the first place, it is essential to consider 
different TPL releases that are conformed to the semantic versioning format.\footnote{\url{https://semver.org/}} A standard version string follows the pattern \emph{X.Y}, in which \emph{X} represents the \emph{major} release and \emph{Y} represents the \emph{minor} one. Sometimes, releases can include a \emph{patch} version \emph{Z}, resulting in the final string \emph{X.Y.Z}. 
We present an explanatory example related to 
\emph{log4j},\footnote{\url{https://logging.apache.org/log4j/}} a widely used Java logging library. When it is upgraded from version \emph{1.2} to version \emph{1.3}, as shown in Listing \ref{lst:v12} and Listing \ref{lst:v13}, respectively, a lot of internal changes happened which need to be carefully documented.\footnote{\url{http://articles.qos.ch/preparingFor13.html}} \revised{As it can be noticed, the main change affects the \texttt{Category} class which is replaced by the \texttt{Logger} class. Furthermore, all the former methods that were used by the deprecated class cause several failures at the source code level. For instance, the \texttt{setPriority} method is replaced by \texttt{setLevel} in the new version.} 


\begin{lstlisting}[caption={log4j version 1.2.}, 
label=lst:v12,captionpos=b,style=JavaStyle,numbers=left,xleftmargin=4em,frame=single,framexleftmargin=4.2em]
Category root = Category.getRoot();
root.debug("hello");

Category cat = Category.getInstance(Some.class);
cat.debug("hello");

cat.setPriority(Priority.INFO);
\end{lstlisting}


\begin{lstlisting}[caption={log4j version 1.3.}, 
label=lst:v13,captionpos=b,style=JavaStyle,numbers=left,xleftmargin=4em,frame=single,framexleftmargin=4.2em]
Logger root = Logger.getRootLogger();
root.debug("hello");

Logger logger = Logger.getLogger(Some.class);
logger.debug("hello");

logger.setLevel(LEVEL.INFO);
\end{lstlisting}


Though this is a very limited use case, it suggests that the code refactoring that takes place during the migration is 
an error-prone activity even for a single minor upgrade, \ie from 
version \emph{1.2} to version \emph{1.3}. Additionally, the complexity 
dramatically grows in the case of a major release as it typically requires extra efforts rather than a minor one which are not welcome by the majority of developers 
\cite{kula_developers_2018}. Considering such context, the reduction of the 
time needed for a single migration step, even a minor one, is expected to 
improve the overall development process.

\begin{figure}[t!]
	\centering
	\includegraphics[width=0.95\linewidth]{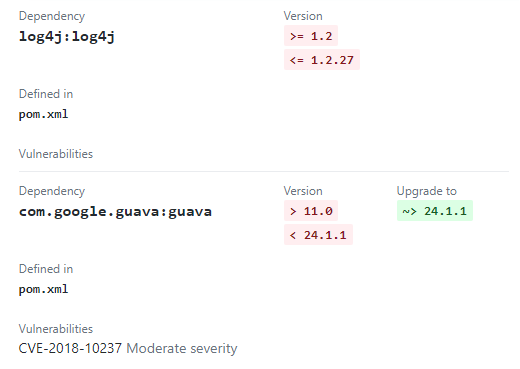}
	\caption{\GH Dependabot alert.}	
	\label{fig:digest}
\end{figure}

Concerning vulnerable dependencies, \GH Dependabot\footnote{\url{https://dependabot.com/blog/github-security-alerts/}} provides weekly security alert digests that highlight possible security issues for outdated dependencies of a repository, 
which can be of different languages, \eg Python, Java, JavaScript.\footnote{\url{https://dependabot.com/\# languages}} An example of a Dependabot report is shown in Fig.~\ref{fig:digest}.

As shown in Fig. \ref{fig:digest}, Dependabot suggests possible TPL upgrades to solve vulnerabilities in the given project. For instance, the \textbf{guava} dependency seems to be outdated, and thus the system automatically suggests jumping to the latest version, \ie \emph{24.1.1}.
 Though this alert can raise awareness 
of this evolution, 
it does not offer any concrete recommendations on how to perform the actual migration steps. In some cases, the bot does not provide any recommended version to update the project, \eg for the \textbf{log4j} dependence. In this respect, we see that there is an urgent need for providing recommendations of the most suitable plan, so as to upgrade the library, as this can significantly reduce the migration effort.

\subsection{Existing techniques} \label{sec:related}

This section reviews some relevant work that copes with the migration problem.

\begin{table}[h]
\centering
\footnotesize
\caption{\revised{Main features of TLPs migration systems.}}
\begin{tabular}{|l | c|  c | c| c| c | c | c | }
\hline
\textbf{System} & \rotatebox[origin=l]{90}{\textbf{Inferring migration}} & \rotatebox[origin=l]{90}{\textbf{Incremental plan}}  & \rotatebox[origin=l]{90}{\textbf{Popularity}}   &  \rotatebox[origin=l]{90}{\textbf{\GH issues}}    & \rotatebox[origin=l]{90}{\textbf{\textbf{Upgrading}}}  &  \rotatebox[origin=l]{90}{\textbf{Replacement}} & \rotatebox[origin=l]{90}{\textbf{Applying Migration}} \\ \hline
Meditor~\cite{xu_meditor_2019} &  \cmark &	\xmark &\cmark  & \xmark & \cmark & \cmark & \cmark   \\ \hline
Apivawe~\cite{hora_apiwave_2015} & \cmark & \xmark & \cmark  & \xmark & \xmark & \cmark & \xmark  \\ \hline
Graph Mining~\cite{teyton_mining_2012}  & \cmark & \xmark & \cmark  & \xmark & \xmark & \cmark & \xmark  \\ \hline
RAPIM~\cite{alrubaye2019learning} & \cmark & \xmark & \cmark & \xmark & \xmark & \cmark & \cmark \\ \hline
Diff-CatchUp~\cite{xing_api-evolution_2007}  & \cmark & \xmark & \cmark  & \xmark & \cmark & \cmark & \xmark  \\ \hline
M$^{3}$~\cite{collie_m3_2020}  & \cmark & \xmark & \xmark  & \xmark & \xmark & \cmark & \cmark  \\ \hline
\rowcolor{mygray}
\textbf{EvoPlan} & \cmark & \cmark & \cmark  & \cmark & \cmark & \xmark & \xmark \color{black}  \\ \hline
\end{tabular}
\label{tab:features}
\end{table}

Meditor \cite{xu_meditor_2019} is a tool aiming to identify migration-related (MR) changes within commits and map them at the level of source code with a syntactic program differencing algorithm. To this end, the tool mines \GH projects searching for MR updates in the \emph{pom.xml} file and check their consistency with the WALA framework.\footnote{\url{https://github.com/wala/WALA}}

Hora and Valente propose Apiwave \cite{hora_apiwave_2015}, a system that excerpts information about libraries' popularity directly from  mined \GH project's history. Afterwards, it can measure the popularity of a certain TLP by considering the import statement removal or addition. 


Teyton \etal~\cite{teyton_mining_2012} propose an approach that discovers migrations among different TLPs and stores them in a graph format. A token-based filter is applied on \emph{pom.xml} files to extract the name and the version of the library from the artifactid tag. The approach evetually exhibits four different visual patterns that consider both ingoing and outgoing edges to highlight the most popular target. 

RAPIM~\cite{alrubaye2019learning} employs a tailored machine learning model to identify and recommend API mappings learned from previously migration changes. Given two TPLs as input, RAPIM extracts valuable method descriptions from their documentation using text engineering techniques and encode them in feature vectors to enable the underpinning machine learning model.

Diff-CatchUp \cite{xing_api-evolution_2007} has been conceived with the aim of proposing usage examples to support the migration of reusable software components. The tool makes use of the UMLDiff algorithm \cite{10.1145/1101908.1101919} to identify all relevant source code refactorings. Then, a heuristic approach is adopted to investigate the design-model of the evolved component and retrieve a customizable ranked list of suggestions. 


%

Collie \etal recently proposed the M$^{3}$ tool \cite{collie_m3_2020} to support a semantic-based migration of C libraries. To this end, the system synthesizes a behavioral model of the input project by relying on the LLVM intermediate representation.\footnote{\url{https://llvm.org/}} Given a pair of source and target TLPs, the tool generates abstract patterns that are used to perform the actual migration.


Table \ref{tab:features} summarizes the features of the above-ment\-ioned approaches by considering the different tasks involved in migration processes by starting with the discovery of possible migration changes up to embedding them directly into the source code as explained below. 

\begin{itemize}
	\item \emph{Inferring migration}: To extract migration-related information, tools can analyze existing projects' artifacts, \ie commits, \emph{pom.xml} file, or tree diff. This is the first step of the whole migration process.
	\item \emph{Incremental plan}: The majority of the existing approaches perform the migration just by considering the latest version of a TLP. This could increase the overall effort needed to perform the actual migration, \ie developers suffer from accumulated technical debt. In contrast, considering a sequence of intermediate migration steps before going to the final one can reduce such refactoring. 
	\item \emph{Popularity}: This is the number of client projects that make use of a certain library. In other words, if a TLP appears in the \emph{pom.xml} file or in the import statement, its popularity is increased. 
	\item \emph{\GH issues}: As an additional criterion, the migration process can include data from \emph{\GH issues} that may include relevant information about TLPs migration. Thus, we consider them as a possible source of migration-related knowledge. 
	\item \emph{Upgrading}: This feature means that the tool supports the upgrading of a TLP from an older version to a newer one. For instance, the  migration described in Section \ref{sec:exp} falls under this class of migration. 
	\item \emph{Replacement}: Differently from upgrading, replacement involves the migration from a library to a different one that exposes the same functionalities. 
	\item \emph{Applying migration}: It represents the final step of the migration process in which the inferred migration changes are actually integrated into the project.
\end{itemize}
%

\subsection{Dimensions to be further explored}
Even though several approaches successfully cope with TPL migration, there are still some development dimensions that need to be further explored. However, providing an exhaustive analysis is out of the scope of this section. Thus, we limit ourselves to identify some of them by carefully investigating the approaches summarized in Table \ref{tab:features}. The elicited dimensions are the following ones:


\begin{itemize}
	\item \emph{D1: Upgrading the same library.}  Almost all of the presented 	
	approaches apart from Meditor, focus on replacing libraries and very few 	
	support the upgrades of already included ones (see columns 	
	\textit{Upgrading} and \textit{Replacement} in Table \ref{tab:features}).
	\item \emph{D2: Varying the migration data sources.}  During the inferring 
	migration phase, strategies to obtain migra\-tion-related data play a 
	fundamental role in the overall process. A crucial challenge should be 
	investigating new sources of information besides the well-known sources 
	\eg  Bug reports, Stack Overflow posts, and \GH issues.
	\item \emph{D3: Aggregating different concepts.} The entire migration 
	process is a complex task and involves notions belonging to different 
	domains. For instance, \GH issues could play a relevant role in the 
	migration process. A recent work \cite{misra_is_2020} shows that
	the more comments are included in the source code, the lesser is 
	the time needed to solve an issue. Neil \etal \cite{neil_mining_2018} extracted 
	security vulnerabilities from issues and bug reports that could affect library dependencies. 
	\item \emph{D4: Identification of the upgrade plan.} Existing approaches 
	identify and apply migrations by taking as input the explicit specification 
	of the target version of the library that has to be upgraded. Providing developers with insights about candidate upgrade plans that might reduce 
	the migration efforts can represent valuable support to the overall upgrade 
	process.
\end{itemize}


In the present work we aim to explore and propose 
solutions for the dimensions \textsc{D1} and \textsc{D4} by providing multiple 
possible upgrade plans given the request of upgrading a given library to target 
a specific target version. Furthermore, we also perform an initial 
investigation on the \textsc{D2} and \textsc{D3} dimensions, relying on \GH 
issues. As it can be seen in Table~\ref{tab:features}, \EP covers five out of  
the seven considered features. In particular, our approach is able to 
\emph{infer migration}, make use of \emph{incremental plan} by considering the 
\emph{popularity} and \emph{issues}, so as to eventually recommend an 
\emph{upgrade plan}. Compared to the existing tools, \EP tackles most of the 
issues previously presented.


	\section{Proposed approach}
	\label{sec:ProposedApproach}
In this paper we propose an approach to support the first phase of the 
migration process, \ie inferring the possible upgrade plans that can satisfy the request of the developer that might want to upgrade a given TPL used in the project under development. 
%
%

Our approach aims at suggesting the most appropriate migration plan by 
taking into consideration two key factors: the \textit{popularity} 
of the upgrade plan and the \textit{availability of discussions} about it.
Popularity means how many clients have performed a given 
upgrade plan, while discussions are \GH issues
that have been open and closed in projects during the migration phase.
%
%
By mining \GH using the dedicated API,\footnote{\url{https://developer.github.com/v3/}} we are able to extract the information required as input for the recommendation 
engine of \EP.

\begin{figure}[t!]
	\centering
	\includegraphics[width=0.90\linewidth]{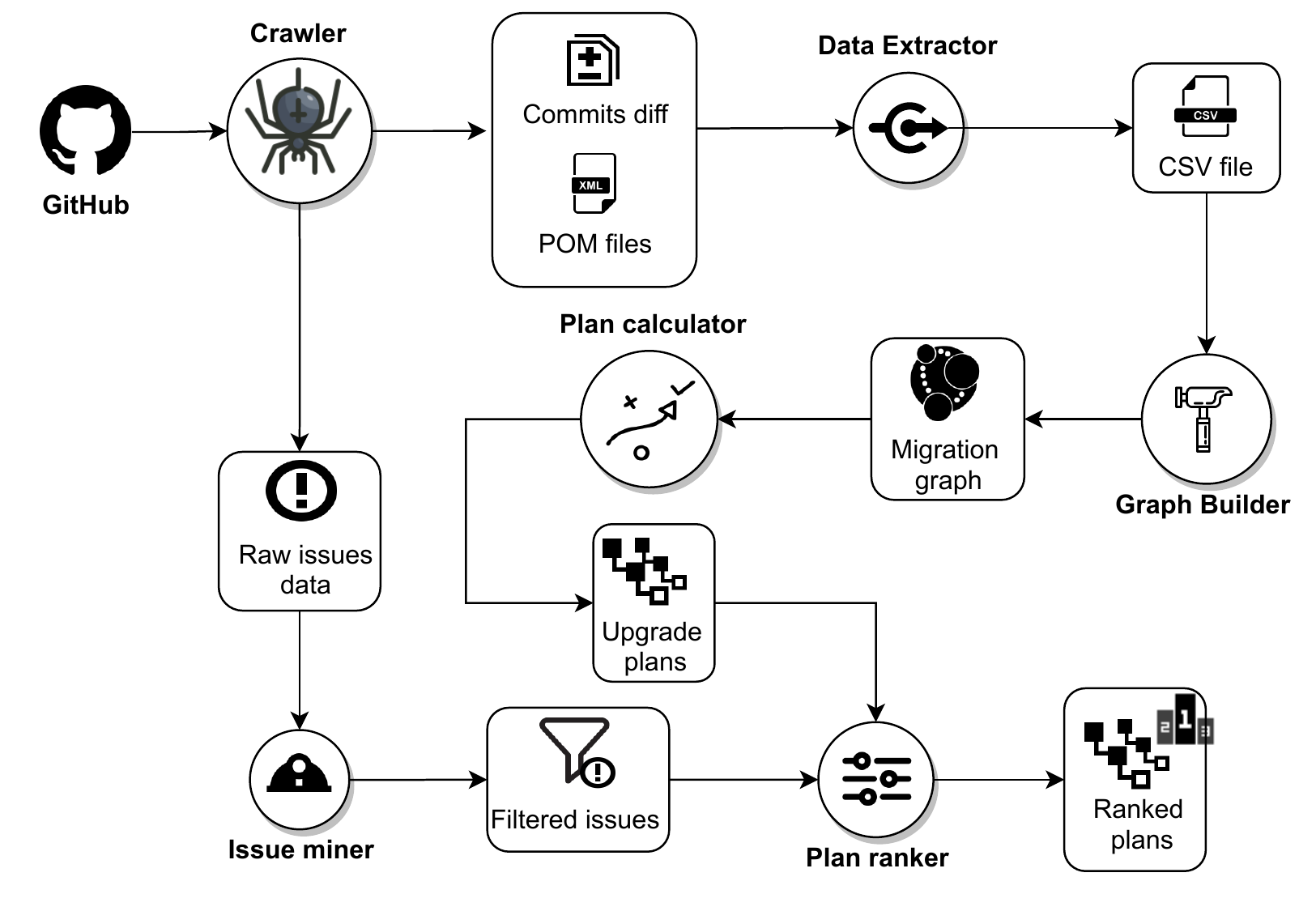}		
	\caption{\EP's architecture.}
	\label{fig:approach}
\end{figure}

The conceived approach is depicted in Fig.~\ref{fig:approach} and consists of six components, \ie \emph{Crawler}, \emph{Data Extractor}, \emph{Graph Builder}, \emph{Issues Miner}, \emph{Plan Calculator} and \emph{Plan Ranker}. With the \emph{Crawler} component, the system retrieves information about \GH repositories and downloads them locally. These repositories are then analyzed by the \emph{Data Extractor} component to excerpt information about commits and history version. Once all the required information has been collected, \emph{Graph Builder} constructs a migration 
graph with multiple wei\-ghts, and \emph{Issues Miner} generates data related to \GH issues. The \emph{Plan Calculator} component relies on the graph to calculate the k-best
paths available. Finally, \emph{Plan Ranker} sorts these paths by considering the number of issues. In the succeeding subsections, we are going to explain in detail the functionalities of each component. 



\subsection{Crawler}  \label{sec:tracker} 
Migration-related information is mined from \GH using the \emph{Crawler} 
component. By means of the \texttt{JGit} library,\footnote{\url{https://www.eclipse.org/jgit/}} \emph{Crawler} downloads a set \textit{P} of \GH projects that have at least one 
\emph{pom.xml} file, which is 
a project file containing the list of all adopted TPLs. In case there are 
multiple \emph{pom.xml} files, they will be analyzed separately to avoid 
information loss. Then, the \emph{Crawler} component analyzes all the 
repository's commits that affect the \emph{pom.xml} to find added and removed 
TPLs. Additionally, raw issue data is obtained and stored in separate files. In 
particular, we count the number of opened and closed issues for each project 
\textit{p} $\in$ \textit{P} in a specific time interval \textit{D}.
The starting point of this interval is identified when a certain version 
\textit{v} of a given library \textit{l} that is added as dependencies of the 
\emph{pom.xml} file in client \textit{C}. A previous study 
\cite{10.1007/978-3-319-26844-622} demonstrates that
the monthly rate of open issues tends to decrease over time. 
Thus, the endpoint of \textit{D} is obtained by considering the first two months of 
development to extract relevant data concerning the considered library \textit{l} without loss of data. \revised{In such a way, the \GH issues that have been opened and closed for each TLP that has been added in \textit{p}, are obtained for further processing phases.}

\subsection{Data Extractor}  \label{sec:dataEx}

\begin{figure}
	\small
		\centering
		\includegraphics[width=\linewidth]{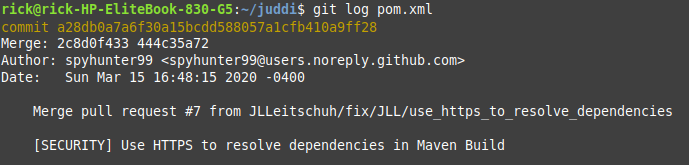} \\ 	
		a) Example of \emph{log} \\
		\includegraphics[width=\linewidth]{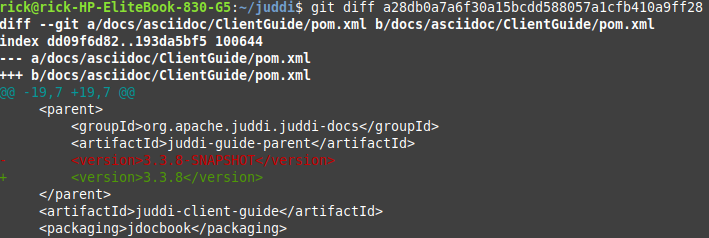}\\
		b) Example of \emph{diff}\\
	\caption{Example of artifacts used by the \emph{Data Extractor} component.}	
	\label{fig:diff-log}	
\end{figure}

In this phase, \revised{data is gathered by means of \texttt{JGit}, and analyzed using different processing steps as follows.}
The first step makes use of the \GH \emph{log} command to retrieve the list of every modification
saved on \GH for a specific file. Furthermore, the command provides the code 
\emph{SHA} for every commit, which allows us to identify it. 
For instance, Fig. \ref{fig:diff-log}.a depicts a 
commit related to a given \emph{pom.xml} file taken as input. The identifier of 
the commit is used to retrieve the list of the corresponding operated changes 
as shown in Fig.  \ref{fig:diff-log}.b. In particular, inside a commit we can 
find a large number of useful information like what was written or removed and 
when. The  \emph{Data Extractor} component focuses on the lines which contain an 
evidence of library changes. In a commit, the added lines are marked with the 
sign '+', whereas the removed ones are marked with '-' (see the green and red lines, respectively shown in Fig.~\ref{fig:diff-log}.b). 
In this way, the evolution of a library is obtained by analyzing the sequence 
of added/removed lines. With this information, \revised{\EP is also able to count} how many 
clients have performed a specific migration. The information retrieved by the 
\emph{Data Extractor} component is stored in a target CSV file, which is taken as input by the subsequent entity of the process as discussed below.

%

%

\subsection{Graph Builder} 
\begin{figure}[t]
	\centering
	\includegraphics[width=1\linewidth]{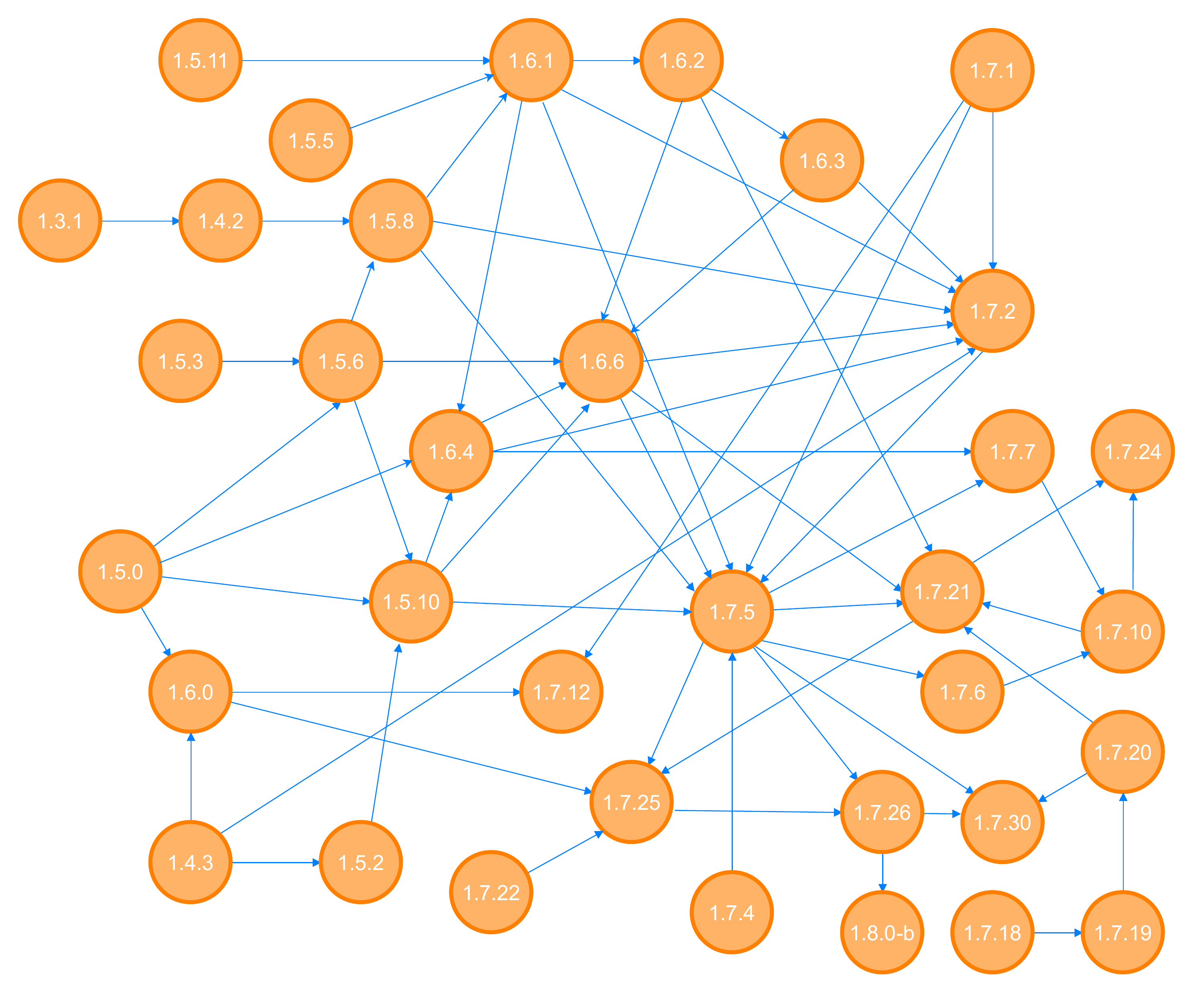}
	\caption{\revised{Migration graph of the \emph{slf4j} library.}}	
	\label{fig:graph}
\end{figure}

This component creates nodes and 
relationships by considering the date and library changes identified in the previous 
phase. To this end, \EP exploits the Cypher query language\footnote{\url{https://neo4j.com/developer/cypher-query-language/}} to store 
data into a \textsc{Neo4j} graph. For instance, we extract from CSV files two pairs library-version \emph{(l,v1)} and \emph{(l,v2)} with signs '-' and '+', respectively. In this way, the component creates an oriented edge from 
\emph{(l,v1)} to \emph{(l,v2)}. Once the first edge is created, any further pairs containing the same library upgrade will be added as an incremented weight on the graph edge.
The date value contained in the CSV record is 
used to avoid duplicated edges or loops. Furthermore, each edge is weighted 
according to the number of clients as described in \textit{Data Extractor} 
phase. That means if we find \emph{w} times the same couple \emph{(l,v1)} to \emph{(l,v2)} (\ie a number of \emph{w} projects have already migrated the library \emph{l} from \emph{v1} to \emph{v2}), the edge will have a weight of \emph{w}.
Thus, the final outcome of 
this component is a migration graph that considers the community's interests as 
the only weight.  For instance, Fig.~\ref{fig:graph} 
represents the extracted migration graph for the \emph{slf4j-api} library. The 
graph contains all the mined version of the library and for each pair the 
corresponding number of clients that have performed the considered upgrade is 
shown. For instance, in Fig.~\ref{fig:graph}  the edge from the version 
\emph{1.6.1} to \emph{1.6.4} is selected, and 14 clients (see the details on 
the bottom) have performed such a migration.

\subsection{Plan Calculator} \label{sec:plan}

Such a component plays a key role in the project. Given a library to be 
upgraded, the starting version, and the target one, \emph{Plan Calculator} 
retrieves
%
%
the k shortest paths 
by using the well-founded \emph{Yen's K-shortest paths 
algorithm} \cite{Yen2007FindingTK} which has been embedded into the 
\textsc{Neo4j} library.
%
As a default heuristic implemented in \EP, the component retrieves all the 
possible paths that maximize the popularity of the steps that can be performed to do the wanted
upgrade. Thus, the \textit{Plan Calculator} component employs the aforementioned weights
which represent the popularity as a criteria for the shortest path algorithm. 
%
%

\begin{figure*}[t]
	\centering
	\includegraphics[width=\linewidth]{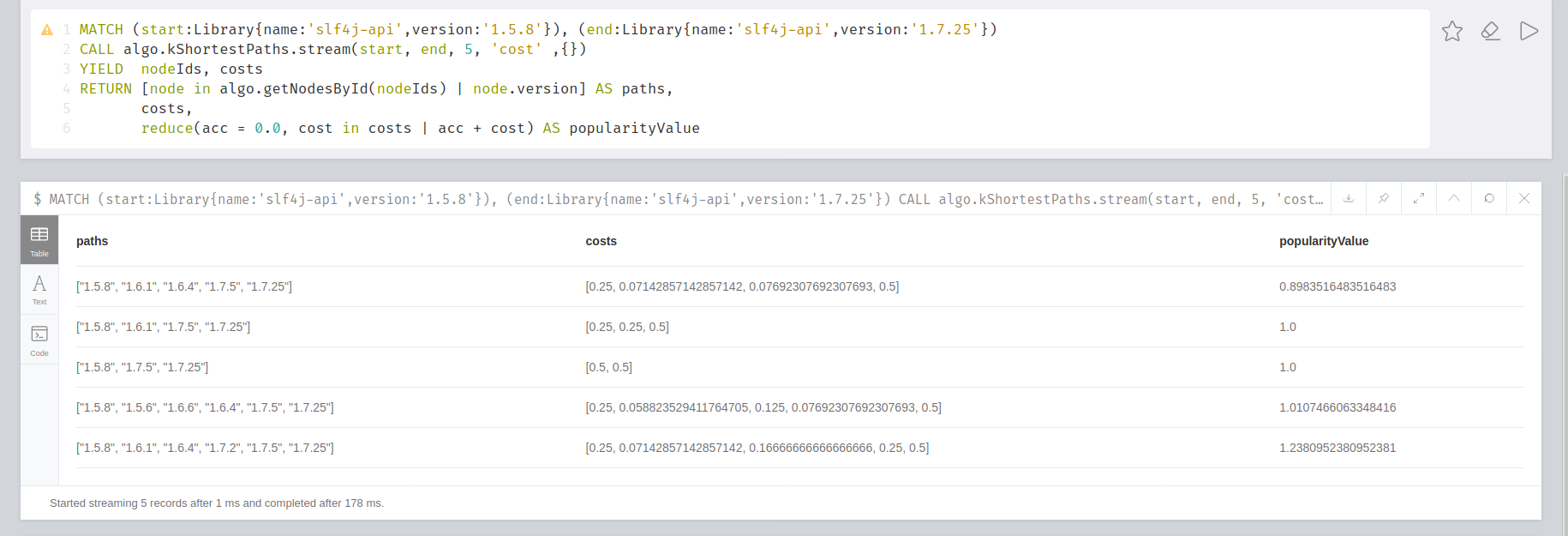}
	\caption{List of \emph{k}-shortest paths for \emph{slf4j}.}
	\label{fig:path}
\end{figure*}

By considering the graph shown in Fig. \ref{fig:graph}, there are several possibilities to upgrade \textit{slf4j} from version \emph{1.5.8} to 
\emph{1.7.25}. By taking into account the available weights, \EP can recommend the ranked list depicted in Fig.~\ref{fig:path}. The first path in the list suggests to follow the steps \emph{1.6.1}, \emph{1.6.4}, and 
\emph{1.7.5} to reach the final version considered in the example, \ie 
\emph{1.7.25}.\footnote{\revised{It is worth noting that the popularity values are disproportionate to the popularity of the corresponding upgrade plans. In the example shown in Fig. \ref{fig:path} the most popular upgrade is the one with popularity value 0.898.}} Such a plan is the one that is performed most by other projects, which rely on \emph{slf4j} and that have already operated the wanted library migration. Thus, such a path is more frequent than directly updating the library to the newest version.

\subsection{Issues Miner} \label{sec:IssuesCalc}

Issues play an important role in project development. For instance, by solving issues, developers contribute to the identification of bugs as well as the enhancement of software quality through feature requests~\cite{liao_exploring_2018}. In the scope of this work, we exploit issues as criteria for ordering upgrade plans. In particular, we rely on the availability of issues that have been opened and closed due to upgrades of given third-party libraries.

\begin{table}[t!]
	\centering
	\caption{Issues information extracted for \emph{commons-io}.}
	\begin{tabular}{|l | c |   c | c | }
		\hline
		\textbf{Version} & \textbf{Open Issues}& \textbf{Closed Issues} & \textbf{Delta}  \\ \hline
		1.0&14&33&19 \\ \hline
		1.3.2&150&420&270 \\ \hline
		1.4&87&408&321 \\ \hline
		2.0&5&10&5 \\ \hline
		2.0.1&133&457&324 \\ \hline
		2.1&129&516&387 \\ \hline
		2.2&67&999&932 \\ \hline
		2.3&5&20&15 \\ \hline
		2.4&939&3,283&2,344 \\ \hline
		2.5&64&918&854 \\ \hline
		2.6&64&548&484 \\ \hline
	\end{tabular}
	\vspace{-.2cm}
	\label{tab:issues}
\end{table}

The \emph{Issue Miner} component is built to aggregate and filter raw issues 
data gathered in the early stage of the process shown in Fig. \ref{fig:approach}. However, due to the internal construction of \textsc{Neo4j}, we cannot directly embed this data as a weight 
on the migration graph's edges. Thus, as shown in Section \ref{sec:tracker}, we 
collect the number of open and closed issues considering a specific time 
window, \ie two months starting from the introduction of a certain TLP in the 
project. Then, this component filters and aggregates the issues data related by 
using Pandas, a widely-used Python library for data mining 
\cite{pandas_pandas_2020}. For instance, Table \ref{tab:issues} shows the mined 
issues related to the \emph{commons-io} library. In particular, for each 
version of the library, the number of issues that have been opened and closed by 
all the analysed clients since they have migrated to that library version is 
shown. \EP can employ the extracted data to enable a ranking function based on \GH issues as discussed in the next section. 

\emph{Issues Miner} works as a stand-alone component, thus it does not impact on the time required by the overall process. In this way, we have an additional source of information that can be used later in the process as a supplementary criterion to choose the ultimate upgrade plan from the ranked list produced by the \textit{Plan Calculator} component.

\subsection{Plan Ranker} \label{sec:PlanRank}
In the final phase, the k-paths produced by the \textit{Plan Calculator} are 
rearranged according to the information about issues. For every 
path, we count the average value of opened/closed issues. A large value means that a certain
path potentially requires less integration effort since there are more closed issues than the opened ones \cite{liao_exploring_2018}, \ie issues 
have been tackled/solved rather than being left untouched. 


Thus, the aim is to order the plans produced by \textit{Plan Calculator} according to the retrieved issues: among the most popular plans we will propose those with the highest issue values.

\begin{table}[h!]
	\caption{An example of the ranking results.}
	\begin{tabular}{|l | p{0.80cm} |  p{0.80cm} | }
		\hline
				\textbf{Proposed Path} & \rotatebox[origin=c]{90}{\textbf{Pop. Value}} & \rotatebox[origin=c]{90}{\textbf{Issues Value}} \\ \hline
		1.5.8, 1.6.1, 1.6.4, 1.6.6, 1.7.5, 1.7.25 &  1.446 &	57   \\ \hline
		\rowcolor{lightgray}
		1.5.8, 1.6.1, 1.6.4, 1.7.5, 1.7.25 &  0.898 &	58   \\ \hline
		1.5.8, 1.7.5, 1.7.25 &  1.0 &	58   \\ \hline
		\rowcolor{Gold}
		1.5.8, 1.6.1, 1.7.5, 1.7.25 &  1.0 &	61   \\ \hline
		1.5.8, 1.6.1, 1.6.4, 1.7.2, 1.7.5, 1.7.25 &  1.238 &	58   \\ \hline
	\end{tabular}
	\label{tab:Ranking}
\end{table}

Table \ref{tab:Ranking} shows an example of the ranking process. There are two highlighted paths, the gray row corresponds to the best result according to the plan popularity only. In fact, the gray highlighted plan is the one with lower popularity value. Meanwhile, the orange row is recommended according to the issues criteria (in this case, the higher the issue value, the better). The path that should be selected is the orange one because it represents the one characterized by the highest activity in terms of opened and closed issue, among the most popular ones. In this way, \EP is 
able to recommend an upgrade plan to migrate from the initial version to the 
desired one by learning from the experience of other projects which have 
already performed similar migrations.

	\section{Evaluation}
	\label{sec:Study}

To the best of our knowledge, there are no replication packages and reusable tools related to the approaches outlined in Section~\ref{sec:Background} that we could use to compare  \EP with them. As a result, it is not possible to compare \EP with any baselines. Thus, we have to conduct an evaluation of \revised{the proposed approach on a real dataset collected from \GH.} 
Section~\ref{sec:ResearchQuestions} presents three research questions, while Section~\ref{sec:Process} describes the evaluation process. Section \ref{sec:dataset} gives a detailed description of the dataset used for the evaluation, and the employed metrics are specified in Section~\ref{sec:metrics}.

\subsection{Research questions} \label{sec:ResearchQuestions}
To study the performance of \EP, we consider the following research questions:
\begin{itemize}
	\item \rqfirst~To answer this question, we conduct experiments following 
	the ten-fold cross-validation methodology~\cite{10.5555/1643031.1643047} on 
	a dataset considering real migration data collected from \GH. Moreover, we 
	compute \emph{Precision}, \emph{Recall}, and \emph{F-measure} by comparing the recommendation outcomes with real migrations as stored in \GH;
	\item \rqsecond \newline We analyze how the number of opened and closed 
	issues 	could affect the migration process. To this end, we compute three 
	different 	statistical coefficients to detect if there exists any 
	correlation among 	the available data.
	\item \rqthird~Besides the recommended migration steps, we are interested in measuring the time of the overall process, including the graph building phase. This aims at ascertaining the feasibility of our approach in practice.
\end{itemize}

\subsection{Overall process} \label{sec:Process}

\begin{figure}[h!]
	\centering
	\includegraphics[width=\linewidth]{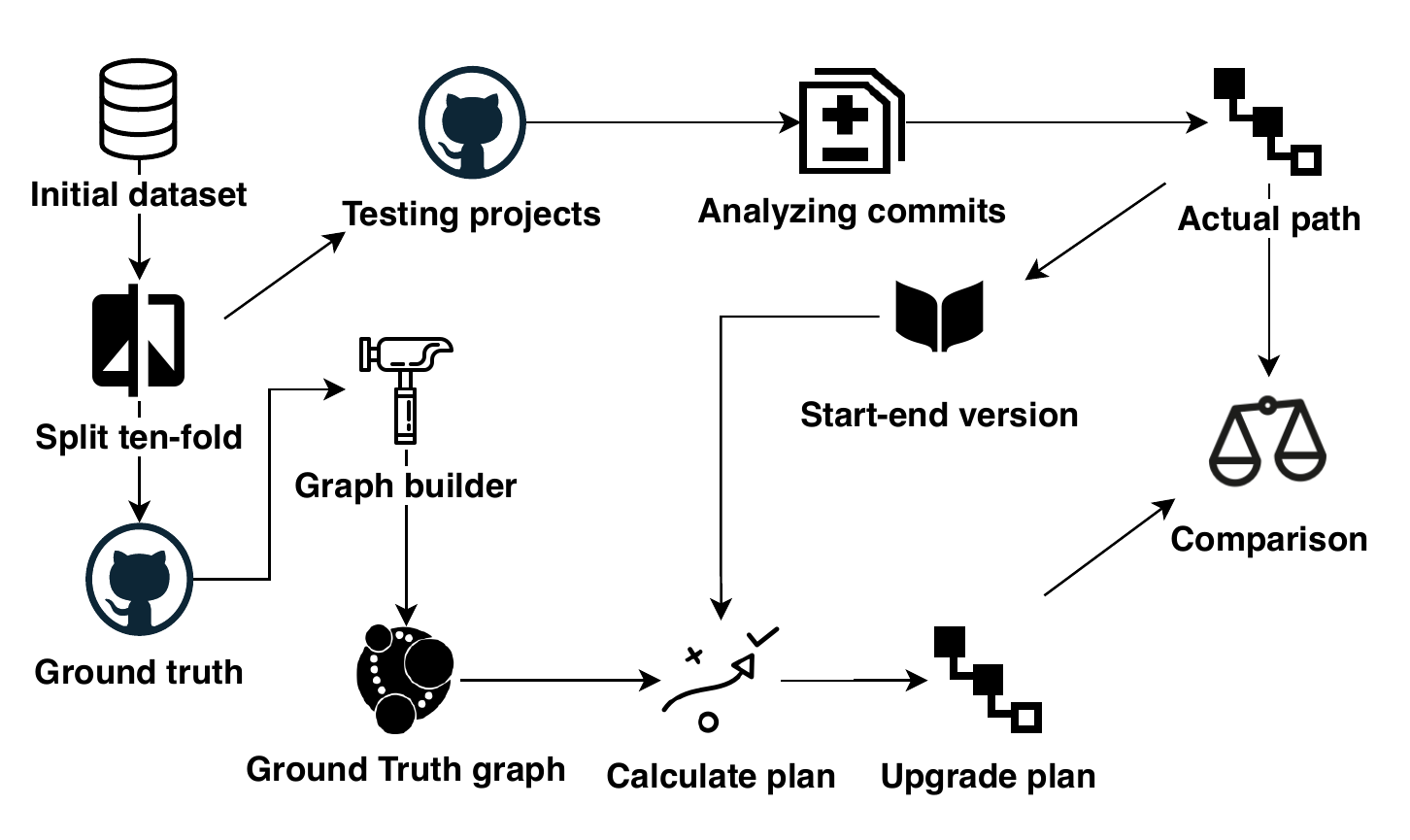}
	\caption{The evaluation process.}
	\label{fig:eval}
\end{figure}

As depicted in Fig.~\ref{fig:eval}, we perform experiments using the ten-fold 
cross-validation methodology on a well-founded dataset coming from an existing 
work~\cite{kula_developers_2018}. 
Given the whole list of $\approx$11,000 projects, we download the entire 
dataset using the \emph{Crawler} component. Then, the dataset is split into testing and 
ground truth projects, \ie 10\% and 90\% of the entire set, respectively, by each round of the 
process. This means that in each round we generate a new migration graph by using the actual 90\% portion. Given a single testing project, the \emph{Analyzing commits} phase is conducted to capture the 
actual upgrade path followed by the repository, as stated in Section \ref{sec:tracker}. 
To build the ground-truth graph, \ie the real migration in \GH, we consider projects not included in the testing ones and calculate 
every possible upgrade plan for each TPLs. 

To aim for a reliable evaluation, we select the starting and the end version of a certain TPL from the actual plan of a testing project. The pair is used to feed the \emph{Plan Calculator} component which in turn retrieves the proposed plan. 
In this respect, by following the two paths we are able to compute the metrics 
to assess the overall performance, namely precision, recall, and F-measure.

\subsection{Data collection} \label{sec:dataset}
We make use of an existing dataset which has been curated by a recent study available on \GH.\footnote{\url{https://bit.ly/2Opd1GH}} 
The rationale behind this selection is the quality of the repositories which were 
collected by applying different filters, \ie removing duplicates, including 
projects with at least one \emph{pom.xml} file, and crawling only well-main\-tained 
and mature projects. 
Table \ref{tab:dataset} summarizes the number of projects and \emph{pom.xml} 
files. The dataset consists of 10,952 \GH repositories, nevertheless we were 
able to download only 9,517 of them, as some have been deleted or moved. 
Starting from these projects, we got a total number of 27,129 \emph{pom.xml} 
files. Among them, we selected only those that did not induce the creation of 
empty elements by the \emph{Data Extractor} component while analyzing 
\textit{logs} and \textit{diffs} as shown in Fig. \ref{fig:diff-log}. The 
filtering process resulted in 13,204 \emph{pom.xml} files. The training set is 
used to create a migration graph to avoid any possible bias. For each round, we 
tested 420 projects, and 3,821 projects are used to build the graph.

\begin{table}[h!]
\centering
\caption{Statistics of the dataset.}
\begin{tabular}{|l | p{1.3cm}|}
\hline
Total number of projects & 10,952  \\ \hline
Number of downloaded projects&  9,517 	\\ \hline
Total number of \emph{pom.xml} files & 27,129  \\ \hline
Number of screened  \emph{pom.xml} files  & 13,204  \\ \hline
\end{tabular}
\label{tab:dataset}
\end{table}

Table \ref{tab:libs} summarizes the set of libraries in the dataset, obtained by employing the \emph{Crawler} module (cf. Section \ref{sec:tracker}). There are seven popular libraries,\footnote{\url{https://mvnrepository.com/popular}} \ie \emph{junit}, \emph{httpclient}, \emph{slf4j}, \emph{log4j}, \emph{commons-io}, \emph{guava}, and \emph{commons-lang3}. 
Among others, \emph{junit} 
has the largest number of migrations, \ie 2,972. Concerning the number of versions, \emph{slf4j} has 71 different versions, being the densest library. Meanwhile, \emph{commons-lang3} is associated with the smallest number of migrations, \ie 162, and \emph{commons-io} is the sparsest library with only 16 versions. The last column shows the number of versions that we could exploit to get the issues. The difference means that no issues data was available for the whole versions dataset.

\begin{table}[h]
\centering
\caption{Number of migrations and versions.}
\begin{tabular}{|l | p{0.80cm} | p{0.80cm} | p{0.80cm} |}
\hline
\textbf{Library} & \rotatebox[origin=l]{90}{\textbf{\# migrations}} & \rotatebox[origin=l]{90}{\textbf{\# versions}} & \rotatebox[origin=l]{90}{\textbf{\# issue vers.}} \\  \hline
\emph{junit} & 2,972 & 30 & 19	\\ \hline
\emph{httpclient} & 218 & 53 & 35	\\ \hline
\emph{slf4j} & 209 & 71 & 26	\\ \hline
\emph{log4j} & 229 & 42 & 19 \\ \hline
\emph{commons-io} & 186 & 16	& 11\\ \hline
\emph{guava} & 627 & 70 & 34	\\ \hline
\emph{commons-lang3} & 162 & 16	& 13\\ \hline
\end{tabular}
\vspace{-.4cm}
\label{tab:libs}
\end{table}

\subsection{Metrics} \label{sec:metrics}

Given a migration path retrieved by \EP, we compare it with the real migration path extracted from a testing project. To this end, we employ 
\emph{Precision}, \emph{Recall}, and \emph{F-measure} (or F$_1$-score)  widely used in the Information Retrieval domain to assess the performance prediction of a system. 
In the first place, we rely on the following definitions:

\begin{itemize}
\item A \textit{true positive} corresponds to the case when the recommended path matches with the actual path extracted from the testing projects; \emph{TP} is the total number of true positives;
\item A \textit{false positive} means that the recommended upgrade plan is not present in the ground-truth paths; \emph{FP} is the total number of false positives; 
\item A \textit{false negative} is the migration steps that should be present in the suggested plan but they are not; \emph{FN} is the total number of false negatives.
\end{itemize}

Considering such definitions, the aforementioned metrics are computed as follows:


\begin{equation} \label{eqn:Precision}
P = \frac{ TP }{TP+ FP}
\end{equation}

\begin{equation} \label{eqn:Recall}
R = \frac{ TP }{TP+FN}
\end{equation}

\begin{equation} \label{eqn:F-Measure}
F-measure = \frac{ 2 \times P \times R}{P + R}
\end{equation}

%
%

\textbf{Rank correlation}: We consider the following coefficients:
\begin{itemize}
	\item \textit{Kendall's tau} 
	measures the strength of dependence between two variables. It is a non-parametric test, \ie it is based on either being distribution-free or having a specified distribution but with the distribution's parameters unspecified. 
	\item \textit{Pearson's correlation} 
	is the most widely used correlation statistic to measure the degree of the relationship between linearly 
	related variables. In particular, this coefficient is suitable when it is possible to draw a regression line between the points of the available data. 
	\item \textit{Spearman's correlation} 
	is a non-parametric test that is used to measure the degree of association between two variables. Differently from Pearson's coefficient, Spearman's correlation index performs better in cases of monotonic relationships. 
\end{itemize}

All the considered coefficients assume values in the range [-1,+1], \ie from perfect negative correlation to perfect positive correlation. The value 0 indicates that between two variables there is no correlation.

In the next section, we explain in detail the experimental results obtained through the evaluation.

%
%
%
%
%


	\section{Experimental results}
	\label{sec:Results}


We report and analyze the obtained results by answering the research questions introduced in the previous section. 


\subsection{\rqfirst}
Table \ref{tab:metrics} reports the average results obtained from the cross-validation evaluation. 
\EP achieves the maximum precision for \emph{commons-io}, \ie 0.90 in all the rounds. The tool also gets a high precision for \emph{junit}, \ie 0.88. Meanwhile, the smallest precision, \ie 0.58 is seen by \emph{httpclient}. Concerning recall, \EP obtains a value of 0.94 and 0.96 for the \emph{junit} and \emph{commons-io} libraries, respectively. In contrast, the tool achieves the worst recall value with \emph{httpclient}, \ie 0.64. Overall by considering the F-Measure score, we see that \EP gets the best and the worst performance by \emph{commons-io} and \emph{httpclient}, respectively.
\vspace{.1cm}

\begin{table}[h!]
\centering
\caption{Precision, Recall, and F-Measure considering popularity.}
\begin{tabular}{|l | p{1.6cm} |  p{1.2cm} | p{1.8cm} |}
\hline
\textbf{Library} & \textbf{Precision} & \textbf{Recall} & \textbf{F-measure} \\ \hline
\emph{junit} &  0.88 &	0.94  &	0.91 \\ \hline
\emph{httpclient} & 0.58 &	0.64 & 0.61 \\ \hline
\emph{slf4j-api} & 0.65  &	0.74 &  0.69  \\ \hline
\emph{log4j} & 0.88&  0.93	& 0.91 \\ \hline
\emph{commons-io} & \textbf{0.90} & \textbf{0.96}  & \textbf{0.94}  \\ \hline
\emph{guava} & 0.60 &  0.73  &	0.65 \\ \hline
\emph{commons-lang3} & 0.66  & 0.67 &	0.65 \\ \hline
\end{tabular}
\label{tab:metrics}
\end{table}


Altogether, we see that there is a substantial difference between the performance obtained by \EP for different libraries. We suppose that this happens due to the availability of the training data. In particular, by carefully investigating each library used in the evaluation, we see that the libraries with the worst results in terms of performance have a few migrations that we can extract from the \emph{pom.xml} on average (cf. Table \ref{tab:libs}). For instance, there are 162 and 209 migrations associated with \emph{commons-lang3} and \emph{slf4j-api}, respectively and \EP gets a low performance on these libraries. Meanwhile, there are 
2,972 migrations for \emph{junit} and \EP gets high precision, recall, and F${_1}$ for this library. It means that less 
data can negatively affect the final recommendations.


Another factor that can influence the conducted evaluation could be the number of versions involved in an upgrade for each library \ie the availability of fewer versions dramatically reduce the migration-related information. This hypothesis is confirmed by the observed values for \emph{log4j} and \emph{junit} that bring better results with 39 and 40 analyzed versions respectively. However, there is an exception with \emph{guava}, \ie \EP yields a mediocre result for the library (F$_{1}$=0.65), though we considered 627 migration paths and 49 different versions. By examining the library, we realized that it has many versions employed in the Android domain as well as abandoned versions. Thus, we attribute the reduction in performance to the lack of decent training data.


\vspace{.2cm}
\begin{tcolorbox}[boxrule=0.86pt,left=0.3em, right=0.3em,top=0.1em, bottom=0.05em]
	\small{\textbf{Answer to RQ$_1$.} \EP is capable of predicting the correct upgrade plan given a real-world migration dataset. Although for some libraries we witness a reduction in the overall performances, the main reason can be found in the lack of migration paths in the original dataset.}
\end{tcolorbox}


\subsection{\rqsecond}
To answer this question we measure the correlation among observed data, \ie the number of clients that perform a certain migration step and the issues delta considering the time interval described in Section \ref{sec:tracker}. 
The number of clients performing migration is defined with the term \textit{popularity} as described in Section \ref{sec:plan}. Meanwhile, as its name suggests, the \textit{delta} is the difference between the number 
of closed issues and the number of open ones. It assumes a positive value when the number of closed issues is greater than the opened ones. In contrast, negative values are observed when open issues exceed the number of closed ones. In other words, deltas characterizes migration steps in terms of closed issues. 

\begin{table}[b!]
	\centering
	\vspace{-.4cm}
	\caption{Correlation coefficients with a $p$-$value < 2.2\mathrm{e}{-16}$.}
	\begin{tabular}{|p{3cm}|p{3cm}|}
		\hline
		\textbf{Metric} & \textbf{Value} \\ \hline
		Kendal's ($\tau$) & 0.458 \\ \hline
		Pearson (r)   & 0.707 \\ \hline
		Spearman ($\rho$)   & 0.616 \\ \hline
	\end{tabular}	
	\label{tab:corr}
\end{table}

The results of the three indexes are shown in Table \ref{tab:corr}. As we can see, all the metrics show a 
positive correlation between the number of clients that perform a certain migration and the corresponding delta issues. In particular, Kendall's tau $\tau$ is equal to 0.458, while Spearman's rank $\rho$ reaches the value of 0.616. The maximum correlation is seen by Pearson's coefficient, \ie r = 0.707.

The strong correlation suggests that given a library, the more clients perform a migration on its versions, the more issues are solved. As it has been shown in a recent work~\cite{liao_exploring_2018}, 
the act of solving issues allows developers to identify bugs and improve code, as well as enhance software quality. Summing up, having a large number of migrated clients can be interpreted as a sign of maturity, 
\ie the evolution among different versions attracts attention 
by developers. 


\vspace{.2cm}

\begin{tcolorbox}[boxrule=0.86pt,left=0.3em, right=0.3em,top=0.1em, bottom=0.05em]
	\small{\textbf{Answer to RQ$_2$.} 
	There is a significant correlation between the upgrade plan popularity and 
	the number of closed issues. This implies that 
	plans to be given highest priority should be those that have the majority 
	of issues solved during the migration.}
\end{tcolorbox}

\subsection{\rqthird}
We measured the average time required for running experiments using a mainstream laptop with the following information: i5-8250U, 1.60GHz Processor, 16GB RAM, and Ubuntu 18.04 as the operating system. Table~\ref{tab:time} summarizes the time for executing the corresponding phases.


\begin{table}[h!]
	\centering
	\vspace{-.3cm}
	\caption{Execution time.}
	\begin{tabular}{|p{3cm}|p{3cm}|}
		\hline
		\textbf{Phase} & \textbf{Time (seconds)} \\ \hline
		Graph building   & 15,120 \\ \hline
		Querying   & 0.11 \\ \hline
		Testing & 145.44 \\ \hline
	\end{tabular}	
	\label{tab:time}
	\vspace{-.2cm}
\end{table}

%


The most time consuming phase is the creation of graph with 15,120 seconds, corresponding to 252 minutes. Meanwhile, the querying phase takes just 0.11 seconds to finish; the testing phase is a bit longer: 145.44 seconds. It is worth noting that the testing consists of the sub-operations that are performed in actual use, \ie opening CSV files, extracting the actual plan, and calculating the shortest path. 
This means that we can get an upgrade plan in less than a second, which is acceptable considering the computational capability of the used laptop.
This suggests that \EP can be deployed in practice to suggest upgrade plans.


\vspace{.2cm}
\begin{tcolorbox}[boxrule=0.86pt,left=0.3em, right=0.3em,top=0.1em, bottom=0.05em]
	\small{\textbf{Answer to RQ$_3$.} The creation of the migration graph is computationally expensive. However, it can be done offline, one time for the whole cycle. \EP is able to deliver a single upgrade plan in a reasonable time window, making it usable in the field.}
\end{tcolorbox}



	\subsection{Threats to validity}
	\label{sec:Threats}
This section discusses possible threats that may affect the proposed approach. 
Threats to \emph{internal validity} could come from the graph building process. 
In particular, the crawler can retrieve inaccurate information from 
\emph{pom.xml} files or \GH commits. To deal with this, we employed a similar 
mining technique used in some related studies presented in Section 
\ref{sec:related}, \ie Meditor, APIwave, aiming to minimize missing data. 
Another possible pitfall lies in downgrade migrations, \ie a client that moves 
from a newer version to an older one. We consider the issue 
as our future work. 

Concerning \emph{external validity}, the main threat is related to the generalizability of the obtained results. We try to mitigate the threat by considering only popular Java libraries. 
Nevertheless, \EP relies on a flexible architecture that can be easily modified to incorporate more TPLs. 
Concerning the employed \GH issues data, they are coarse-grained, \ie we can have a huge number of issues that do not have a strong tie with the examined TLPs. We addressed this issue in the paper by considering the ratio of the delta instead of absolute numbers.  
Concerning the supported data sources, \EP employs \emph{Maven} and \GH to mine migration histories and retrieve issues, respectively. Thus, currently, upgrade plans can be recommended for projects that rely on these two technologies. However, the architecture of \EP has been designed in a way that supporting additional data sources would mean operating localized extensions in the \textit{Crawler}, \textit{Data Extractor}, and \textit{Issue Miner} components without compromising the validity of the whole architecture. 

Finally, threats to \emph{construct validity} concern the ten-fold cross-validation shown in Section \ref{sec:Process}. Even though this technique is used mostly in the machine learning domain, we mitigate any possible incorrect values by considering a different ground-truth graph for each evaluation round. Additionally, the usage of \GH issues could be seen as a possible threat. We mitigate this aspect by using such information as post-processing to reduce possible negative impacts on the recommended items, \ie ranking the retrieved upgrade plans according to the total amount of issues.

	\section{Related work}
	\label{sec:RelatedWorks}
A plethora of studies highlights different issues related to the TLPs migration problem. Dig and Johnson \cite{dig_role_2005} demonstrate the role of code refactorings as the principal origin of \emph{breaking changes}, \ie failures caused by a library upgrade from an older version to a newer one. 
\emph{Binary incompatibilities (BIs)} happen when the application is no longer compilable after migration~\cite{cossette_seeking_2012}. 
The Clirr tool has been used to detect the entities that cause incompatibilities by analyzing the JAR files of the testing project.
By evaluating six different recommendation techniques that are typically used to fix BIs,  this study exhibits that they were capable of resolving only 20\% of them. 

A recent work \cite{kula_developers_2018} attracts the community attention over the \emph{migration awareness} problem. By conducting a user study, the two main migration awareness mechanisms have been evaluated, \ie security advisories and new releases announcement. In this respect, the results show that the majority of the software systems rarely update the older but reliable libraries and security advisories provide incomplete solutions to the developers.

Alrubaye \etal~\cite{alrubaye_how_nodate} conducted an empirical study to highlight the benefits of the migration process over software quality measured by the three standard metrics used in the domain, \ie coupling, cohesion, and complexity. By relying on a dataset composed of nine different libraries and 57,447 Java projects, statistical tests have been carried on relevant migration data. The results confirm that the migration process improves the code quality in terms of the mentioned metrics. 

The problem of \emph{Technical debt} has been studied in both 
academia 
and industry~\cite{avgeriou_et_al:DR:2016:6693}, 
and it is related to 
``immature'' code 
sent to production~\cite{10.1145/157710.157715}. Although this approach is used to achieve immediate results, it could lead to future issues after a certain period. To solve this, technical debt can be repaid through code refactorings by carrying out a cost-benefit analysis. Lavazza \etal \cite{lavazza_technical_2018} propose the usage of technical debt as an external software quality attribute of a project. Furthermore, technical debt can affect software evolution and maintainability by introducing defects that are difficult to fix. 

Sawant and Bacchelli \cite{sawant_fine-grape_2017} investigate \emph{API features usages} over different TLPs releases by mining 20,263 projects and collect 1,482,726 method invocations belonging to five different libraries. Using the proposed tool fine-GRAPE, two case studies have been conducted considering two aspects, \ie the number of migrations towards newer versions and the usages of API features. The results 
confirm that developers tend not to update their libraries. More interesting, the second study shows that a low percentage of API features are actually used in the examined projects.   


	\section{Conclusion and future work}
	\label{sec:Conclusion}


The migration of TPLs during the development of a software project plays an 
important role in the whole development cycle. Even though some tools are 
already in place to solve the issue, different challenges are still opened, \ie 
reducing the effort during the migration steps or the need to consider heterogeneous data 
sources to name a few. We proposed \EP, a novel approach to support the 
upgrading of TPLs by considering miscellaneous software artifacts. By 
envisioning different components, our tool is capable of extracting relevant 
migration data and encoding it in a flexible graph-based representation. Such a 
migration graph is used to retrieve multiple upgrade plans considering the 
popularity as the main rationale. They are eventually ranked by exploiting the 
\GH issues data to possibly minimize the  effort that is required by the 
developer to select one of the candidate upgrade plans. A feasibility study 
shows that the results are promising, with respect to both effectiveness and 
efficiency. 

As future work, we plan to incorporate additional concepts in the 
migration graph, \ie TLPs documentation, Stack Overflow posts, and issues sentiment 
analysis.  We believe that such additional data allows \EP to better capture the 
migration paths performed by clients. Moreover, we can consider a larger 
testing dataset to improve the coverage of the recommendation items, \ie 
provide upgrade plans for more TLPs.

	\begin{acknowledgements}
		The research described in this paper has been partially supported by the AIDOaRT Project, which has received funding from 
		the European Union's H2020-ECSEL-2020, Federal Ministry of Education, Science and Research, Grant Agreement n$^{\circ}$101007350
	\end{acknowledgements}

	%
	%

	\bibliographystyle{spmpsci}      

\bibliography{main}

\begin{thebibliography}{10}
\providecommand{\url}[1]{{#1}}
\providecommand{\urlprefix}{URL }
\expandafter\ifx\csname urlstyle\endcsname\relax
  \providecommand{\doi}[1]{DOI~\discretionary{}{}{}#1}\else
  \providecommand{\doi}{DOI~\discretionary{}{}{}\begingroup
  \urlstyle{rm}\Url}\fi

\bibitem{alrubaye_how_nodate}
Alrubaye, H., Alshoaibi, D., Alomar, E., Mkaouer, M.W., Ouni, A.: How does
  library migration impact software quality and comprehension? an empirical
  study.
\newblock In: S.~Ben~Sassi, S.~Ducasse, H.~Mili (eds.) Reuse in Emerging
  Software Engineering Practices, pp. 245--260. Springer International
  Publishing, Cham (2020)

\bibitem{alrubaye2019learning}
Alrubaye, H., Mkaouer, M.W., Khokhlov, I., Reznik, L., Ouni, A., Mcgoff, J.:
  Learning to recommend third-party library migration opportunities at the api
  level.
\newblock Applied Soft Computing \textbf{90}, 106--140 (2020)

\bibitem{avgeriou_et_al:DR:2016:6693}
Avgeriou, P., Kruchten, P., Ozkaya, I., Seaman, C.: {Managing Technical Debt in
  Software Engineering (Dagstuhl Seminar 16162)}.
\newblock Dagstuhl Reports \textbf{6}(4), 110--138 (2016).
\newblock \doi{10.4230/DagRep.6.4.110}

\bibitem{collie_m3_2020}
Collie, B., Ginsbach, P., Woodruff, J., Rajan, A., O'Boyle, M.F.: M3: Semantic
  api migrations.
\newblock In: 2020 35th IEEE/ACM International Conference on Automated Software
  Engineering (ASE), pp. 90--102 (2020)

\bibitem{cossette_seeking_2012}
Cossette, B.E., Walker, R.J.: Seeking the ground truth: a retroactive study on
  the evolution and migration of software libraries.
\newblock In: Procs. of the {ACM} {SIGSOFT} 20th {Int.} {Symposium} on the
  {Foundations} of {Software} {Engineering} - {FSE} '12, p.~1. Cary, North
  Carolina (2012).
\newblock \doi{10.1145/2393596.2393661}

\bibitem{10.1145/157710.157715}
Cunningham, W.: The wycash portfolio management system.
\newblock SIGPLAN OOPS Mess. \textbf{4}(2), 29–30 (1992).
\newblock \doi{10.1145/157710.157715}.
\newblock
  \urlprefix\url{https://doi-org.univaq.clas.cineca.it/10.1145/157710.157715}

\bibitem{10.1145/3382494.3410690}
Di~Rocco, J., Di~Ruscio, D., Di~Sipio, C., Nguyen, P., Rubei, R.: {TopFilter:
  An Approach to Recommend Relevant GitHub Topics}.
\newblock In: Proceedings of the 14th ACM / IEEE International Symposium on
  Empirical Software Engineering and Measurement (ESEM), ESEM '20. Association
  for Computing Machinery, New York, NY, USA (2020).
\newblock \doi{10.1145/3382494.3410690}

\bibitem{di_rocco_development_2021}
Di~Rocco, J., Di~Ruscio, D., Di~Sipio, C., Nguyen, P.T., Rubei, R.: Development
  of recommendation systems for software engineering: the {CROSSMINER}
  experience \textbf{26}(4), 69.
\newblock \doi{10.1007/s10664-021-09963-7}.
\newblock \urlprefix\url{https://doi.org/10.1007/s10664-021-09963-7}

\bibitem{10.1145/3383219.3383227}
Di~Sipio, C., Rubei, R., {Di Ruscio}, D., Nguyen, P.T.: {Using a Multinomial
  Na\"ive Bayesian (MNB) Network to Automatically Recommend Topics for GitHub
  Repositories}.
\newblock In: Proceedings of the 24th International Conference on Evaluation
  and Assessment in Software Engineering, EASE2020, Trondheim, Norway, April
  15-17, 2020, EASE’20, pp. 24--34. {ACM} (2020).
\newblock \doi{10.1145/3383219.3383227}

\bibitem{dig_role_2005}
Dig, D., Johnson, R.: The role of refactorings in {API} evolution.
\newblock In: 21st {IEEE} {Int.} {Conf.} on {Software} {Maintenance}
  ({ICSM}'05), pp. 389--398 (2005).
\newblock \doi{10.1109/ICSM.2005.90}

\bibitem{hora_apiwave_2015}
Hora, A., Valente, M.T.: Apiwave: {Keeping} track of {API} popularity and
  migration.
\newblock In: 2015 {IEEE} {Int.} {Conf.} on {Software} {Maintenance} and
  {Evolution} ({ICSME}), pp. 321--323 (2015).
\newblock \doi{10.1109/ICSM.2015.7332478}

\bibitem{10.1007/978-3-319-26844-622}
Kikas, R., Dumas, M., Pfahl, D.: Issue dynamics in github projects.
\newblock In: Proceedings of the 16th International Conference on
  Product-Focused Software Process Improvement - Volume 9459, PROFES 2015, p.
  295–310. Springer-Verlag, Berlin, Heidelberg (2015).
\newblock \doi{10.1007/978-3-319-26844-6_22}.
\newblock \urlprefix\url{https://doi.org/10.1007/978-3-319-26844-6_22}

\bibitem{10.5555/1643031.1643047}
Kohavi, R.: A study of cross-validation and bootstrap for accuracy estimation
  and model selection.
\newblock In: Proceedings of the 14th International Joint Conference on
  Artificial Intelligence - Volume 2, IJCAI’95, p. 1137–1143. Morgan
  Kaufmann Publishers Inc., San Francisco, CA, USA (1995)

\bibitem{kula_developers_2018}
Kula, R.G., German, D.M., Ouni, A., Ishio, T., Inoue, K.: Do developers update
  their library dependencies?: {An} empirical study on the impact of security
  advisories on library migration.
\newblock Empirical Software Engineering \textbf{23}(1), 384--417 (2018).
\newblock \doi{10.1007/s10664-017-9521-5}

\bibitem{lavazza_technical_2018}
Lavazza, L., Morasca, S., Tosi, D.: Technical debt as an external software
  attribute.
\newblock In: Proceedings of the 2018 {International} {Conference} on
  {Technical} {Debt} - {TechDebt} '18, pp. 21--30. ACM Press, Gothenburg,
  Sweden (2018).
\newblock \doi{10.1145/3194164.3194168}.
\newblock \urlprefix\url{http://dl.acm.org/citation.cfm?doid=3194164.3194168}

\bibitem{liao_exploring_2018}
Liao, Z., He, D., Chen, Z., Fan, X., Zhang, Y., Liu, S.: Exploring the
  {Characteristics} of {Issue}-{Related} {Behaviors} in {GitHub} {Using}
  {Visualization} {Techniques}.
\newblock IEEE Access \textbf{6}, 24003--24015 (2018).
\newblock \doi{10.1109/ACCESS.2018.2810295}.
\newblock Conference Name: IEEE Access

\bibitem{misra_is_2020}
Misra, V., Reddy, J.S.K., Chimalakonda, S.: Is there a correlation between code
  comments and issues?: an exploratory study.
\newblock In: Proceedings of the 35th {Annual} {ACM} {Symposium} on {Applied}
  {Computing}, pp. 110--117. ACM, Brno Czech Republic (2020).
\newblock \doi{10.1145/3341105.3374009}.
\newblock \urlprefix\url{https://dl.acm.org/doi/10.1145/3341105.3374009}

\bibitem{neil_mining_2018}
Neil, L., Mittal, S., Joshi, A.: Mining {Threat} {Intelligence} about
  {Open}-{Source} {Projects} and {Libraries} from {Code} {Repository} {Issues}
  and {Bug} {Reports}.
\newblock In: 2018 {IEEE} {International} {Conference} on {Intelligence} and
  {Security} {Informatics} ({ISI}), pp. 7--12 (2018).
\newblock \doi{10.1109/ISI.2018.8587375}

\bibitem{Nguyen:2019:JSS:CrossRec}
Nguyen, P.T., {Di Rocco}, J., {Di Ruscio}, D., {Di Penta}, M.: {CrossRec:
  Supporting Software Developers by Recommending Third-party Libraries}.
\newblock Journal of Systems and Software p. 110460 (2019).
\newblock \doi{https://doi.org/10.1016/j.jss.2019.110460}.
\newblock
  \urlprefix\url{http://www.sciencedirect.com/science/article/pii/S0164121219302341}

\bibitem{Nguyen:2019:FRS:3339505.3339636}
Nguyen, P.T., Di~Rocco, J., Di~Ruscio, D., Ochoa, L., Degueule, T., Di~Penta,
  M.: {FOCUS}: {A} recommender system for mining {API} function calls and usage
  patterns.
\newblock In: Proceedings of the 41st international conference on software
  engineering, {ICSE} '19, pp. 1050--1060. IEEE Press, Piscataway, NJ, USA
  (2019)

\bibitem{pandas_pandas_2020}
Pandas: pandas documentation — pandas 1.1.3 documentation (2020).
\newblock \urlprefix\url{https://pandas.pydata.org/docs/}

\bibitem{ponzanelli_prompter:_2016}
Ponzanelli, L., Bavota, G., Di~Penta, M., Oliveto, R., Lanza, M.: Prompter:
  {Turning} the {IDE} into a self-confident programming assistant.
\newblock Empirical Software Engineering \textbf{21}(5), 2190--2231 (2016).
\newblock \doi{10.1007/s10664-015-9397-1}.
\newblock \urlprefix\url{http://link.springer.com/10.1007/s10664-015-9397-1}

\bibitem{robillard_recommendation_2014}
Robillard, M.P., Maalej, W., Walker, R.J., Zimmermann, T. (eds.):
  Recommendation {Systems} in {Software} {Engineering}.
\newblock Berlin, Heidelberg (2014).
\newblock \doi{10.1007/978-3-642-45135-5}

\bibitem{RUBEI2020106367}
Rubei, R., {Di Sipio}, C., Nguyen, P.T., {Di Rocco}, J., {Di Ruscio}, D.:
  {PostFinder: Mining Stack Overflow posts to support software developers}.
\newblock Information and Software Technology \textbf{127}, 106367 (2020).
\newblock \doi{https://doi.org/10.1016/j.infsof.2020.106367}.
\newblock
  \urlprefix\url{http://www.sciencedirect.com/science/article/pii/S0950584920301361}

\bibitem{sawant_fine-grape_2017}
Sawant, A.A., Bacchelli, A.: fine-{GRAPE}: fine-grained {APi} usage extractor
  – an approach and dataset to investigate {API} usage.
\newblock Empirical Software Engineering \textbf{22}(3), 1348--1371 (2017).
\newblock \doi{10.1007/s10664-016-9444-6}.
\newblock \urlprefix\url{https://doi.org/10.1007/s10664-016-9444-6}

\bibitem{teyton_mining_2012}
Teyton, C., Falleri, J.R., Blanc, X.: Mining {Library} {Migration} {Graphs}.
\newblock In: 2012 19th {Working} {Conf.} on {Reverse} {Engineering}, pp.
  289--298 (2012).
\newblock \doi{10.1109/WCRE.2012.38}

\bibitem{7884616}
{Xavier}, L., {Brito}, A., {Hora}, A., {Valente}, M.T.: Historical and impact
  analysis of api breaking changes: A large-scale study.
\newblock In: 2017 IEEE 24th Int. Conf. on Software Analysis, Evolution and
  Reengineering (SANER), pp. 138--147 (2017)

\bibitem{10.1145/1101908.1101919}
Xing, Z., Stroulia, E.: Umldiff: An algorithm for object-oriented design
  differencing.
\newblock In: Proceedings of the 20th IEEE/ACM International Conference on
  Automated Software Engineering, ASE '05, p. 54–65. Association for
  Computing Machinery, New York, NY, USA (2005).
\newblock \doi{10.1145/1101908.1101919}.
\newblock \urlprefix\url{https://doi.org/10.1145/1101908.1101919}

\bibitem{xing_api-evolution_2007}
Xing, Z., Stroulia, E.: {API}-{Evolution} {Support} with {Diff}-{CatchUp}.
\newblock IEEE Transactions on Software Engineering \textbf{33}(12), 818--836
  (2007).
\newblock \doi{10.1109/TSE.2007.70747}

\bibitem{xu_meditor_2019}
Xu, S., Dong, Z., Meng, N.: Meditor: {Inference} and {Application} of {API}
  {Migration} {Edits}.
\newblock In: 2019 {IEEE}/{ACM} 27th {Int.} {Conf.} on {Program}
  {Comprehension} ({ICPC}), pp. 335--346 (2019).
\newblock \doi{10.1109/ICPC.2019.00052}

\bibitem{Yen2007FindingTK}
Yen, J.Y., YENt, J.Y.: Finding the k shortest loopless paths in a network
  (2007)

\end{thebibliography}
\end{document}